# Enhancing Electrical Properties of Selectively Grown In-Plane InAs Nanowires using InGaAs Buffer and Capping Layers

Pradip Adhikari[1], Anjali Rathore[1], Dayrl P Briggs[2], Srijanto R Bernadeta[2], Joon Sue Lee[1,*]

[1]Department of Physics and Astronomy, University of Tennessee, Knoxville, TN 37996, USA

[2]Center for Nanophase Materials Sciences, Oak Ridge National Laboratory, Oak Ridge, TN, 37831, USA

*Corresponding author e-mail: jslee@utk.edu

**Abstract:**

In-plane semiconductor nanowires with complex branched geometries, prepared via selective area growth (SAG), offer a versatile platform for advanced electronics, optoelectronics, and quantum devices. However, defects and disorder at the interfaces and top surfaces of the nanowires can significantly degrade electrical properties. One effective method to mitigate these issues is the incorporation of buffer and capping layers with close lattice matching. In this work, we utilized InGaAs as buffer and capping layers for SAG InAs nanowires after expanding the growth selectivity window for InGaAs in the presence of atomic hydrogen. Hall measurements on InAs nanowires, with and without InGaAs buffer and/or capping layers, revealed that incorporating closely lattice-matched InGaAs buffer and capping layers nearly tripled the electron mobility and doubled the phase coherence length compared to nanowires without these layers. These findings highlight that the use of InGaAs buffer and capping layers is a crucial strategy for significantly enhancing the quality of InAs nanowires, unlocking their full potential for high performance electronics and quantum devices.

## 1. Introduction

III-V semiconductor nanowires are promising material platforms for fundamental research in quantum coherence, light-matter interaction, and topological states, as well as for applications in electronics, photonics and quantum information technologies, owing to their tunable physical and electrical properties and versatility in different growth modes [1–7]. Nanostructure devices can be fabricated using two main approaches: top-down, which involves etching and lithography on bulk materials and thin films, and bottom-up, which employs confined growth techniques, including selective area growth (SAG). SAG is a localized growth method in which the substrate is lithographically patterned with an amorphous mask, allowing growth only on pre-defined patterns under suitable conditions. SAG of in-plane nanowires is preferred for its scalability, ability to create complex nanowire geometries, and compatibility with commonly used growth methods such as molecular beam epitaxy (MBE) [8–13], chemical beam epitaxy (CBE), [14,15] and metal-organic chemical vapor deposition (MOCVD) [16–18].

Despite the promising aspects of SAG in-plane III-V nanowires, one of the major challenges is the formation of defects and disorder at their interfaces and surfaces. Processes involved in substrate preparation and native oxide removal can introduce disorder on the substrate surface, and lattice mismatch between the substrate and nanowire induces strain-related defects, such as misfit dislocations and stacking faults. Other sources of defects and disorder include dangling bonds, damage to the nanowire surface during post-growth fabrication, and the formation of native oxides, all of which degrade device performance. To address such issues in epitaxial growths of III-V as well as non-III-V semiconductors, efforts have been made to grow materials on lattice-matched substrates [13] and to incorporate buffer [8,19,20] and capping layers [8,13,20]. Adding buffer layers with lattice parameters closer to those of the nanowire material is advantageous as they bury the disordered surface of the substrate and reduce defects stemming from the lattice mismatch between the substrate and nanowire, whereas capping layers may prevent nanowire degradation by protecting it from damage during post-growth processing and by avoiding the formation of an oxide layer on the top layers of the conduction path of the nanowire. A significant increase in electron mobility due to the addition of suitable capping layers (top barriers) has been demonstrated for the case of InAs two-dimensional electron gas systems [21–24]. Additionally, in quantum devices based on superconductor-semiconductor hybrid systems, capping layers prevent metallization of the semiconductor and can be used as a control knob to modulate the coupling between the superconductor and semiconductor [5,25,26].

Among III-V semiconductors, SAG of in-plane InAs has gained significant attention as one-dimensional InAs channels can be readily proximitized by superconductors to enable quantum device applications, thanks to their high electron mobility, strong spin-orbit coupling, and narrow bandgap [27]. To improve the performance of SAG InAs nanowires, buffer and capping layers of $In_xGa_{1-x}As$, whose lattice constant closely matches to that of InAs, can be employed. However, while adding $In_xGa_{1-x}As$ as buffer and capping layers in the SAG of InAs, it is essential to establish appropriate growth conditions for SAG $In_xGa_{1-x}As$. Prior research indicates that the selectivity windows for SAG of InAs and GaAs have minimal overlap [28], resulting in a narrow growth window for $In_xGa_{1-x}As$. This often necessitates higher growth temperatures for SAG of InGaAs, however, this leads to the unwanted Ga intermixing to InAs channels [8]. To address this, our work focuses on widening the selectivity growth window for GaAs, consequently InGaAs, by using atomic hydrogen (H) during SAG of InAs with InGaAs buffer and capping layers to enhance electrical properties, such as electron mobility and phase coherence length. To reveal the effect of buffer and capping layers on electrical properties, four sets of SAG InAs nanowires were prepared: (a) InAs, (b) InAs with an InGaAs capping layer, (c) InAs with an InGaAs buffer layer, and (d) InAs with both InGaAs buffer and capping layers. Pre-patterned Hall-bar devices were utilized to characterize electrical properties through electrical transport measurements. Enhanced electrical properties, including higher electron mobility and longer phase coherence length, were obtained by adding buffer and/or capping layers. The Hall mobility for (d) InAs with both buffer and capping layers was almost three times higher than that of (a) InAs without buffer and capping layers. By improving the fundamental understanding of the growth process of SAG nanowires and enhancing electrical properties by adding buffer and capping layers, our work opens possibilities for more efficient and high-performance electronic, optoelectronic, and quantum devices based on SAG III-V semiconductor nanowires.

## 2. Results and Discussion

### A. Selectivity window for SAG InAs, GaAs, and InGaAs

The SAG method utilizes a substrate with a lithographically patterned mask, enabling selective nucleation of the target material exclusively through the mask openings. The conditions that optimize this selective growth are referred to as the "selectivity window." However, achieving growth selectivity depends on several factors, such as mask material, growth material, and growth conditions. During SAG of III-V nanowires by MBE, four growth modes are observed [28]: (i) parasitic nucleation of group-III droplets on the mask concurrent with growth inside the openings, (ii) parasitic nucleation of III-V crystallites on the mask along with growth inside the openings, (iii) selective growth exclusively inside the mask openings, and (iv) no growth on the mask and in the openings. In typical SAG by MBE, under group-V rich conditions, nucleation of group-III droplets is rare [29], leaving three feasible growth modes. Since the primary factor governing selectivity is the desorption of group-III adatoms from the mask and III-V substrate [28], we focus on determining the selectivity window for group III flux under group-V-rich conditions.

Figure 1 shows the selectivity window for SAG of InAs, GaAs, and InGaAs with a silicon nitride ($SiN_x$) mask along with scanning electron microscopy (SEM) images of nanowires grown in selective growth regions. The red circles and blue diamonds in Fig. 1(a) and Fig. 1(b) represent the transition points between the selective growth mode and the parasitic growth mode (upper limit) while the black squares represent the transition points between the selective growth mode and the no-growth mode (lower limit), obtained by monitoring *in-situ* reflection high energy electron diffraction (RHEED) patterns. The details of how we obtained the selectivity window are explained in Supplemental Material S1.

The selectivity window of InAs and GaAs can be extended to obtain the selectivity window for the ternary alloy $In_xGa_{1-x}As$. The region in which both InAs and GaAs can be selectively grown defines the selective growth region for $In_xGa_{1-x}As$. However, there is a narrow overlap between the InAs and GaAs selectivity windows, giving rise to a limited parameter space for the selective growth of $In_xGa_{1-x}As$. Attempts have been made to widen the selectivity window of III-V materials by employing growth techniques such as metal-modulated epitaxy [30] and migration-enhanced epitaxy [31], however these techniques alter the morphology and transport characteristics of the grown layers [32,33]. Another approach to widen the selectivity window without altering morphology involves using atomic H during growth. Atomic H has been employed in the SAG of GaAs [34], GaSb [35], and InGaAs [36] at comparatively lower temperatures. In this work, we obtained a widened selectivity window for the GaAs and, consequently, $In_xGa_{1-x}As$ using atomic H during SAG. Use of atomic H lowers the desorption temperature of Ga from the amorphous mask by forming volatile compounds like gallium hydrides. We repeated the same process as described in supplemental information S1B to obtain the upper limit of Ga flux in the presence of atomic H. The shifted upper limit for GaAs, and hence $In_xGa_{1-x}As$, is indicated by the blue line in Figs. 1(b) and 1(c). The SEM images of $In_{0.75}Ga_{0.25}As$ nanowires, shown in the inset of Fig. 1(c), grown with and without atomic H under the same growth conditions, also indicate improved selectivity with the use of atomic H. However, no change was observed in the selectivity window of InAs with the use of atomic H.

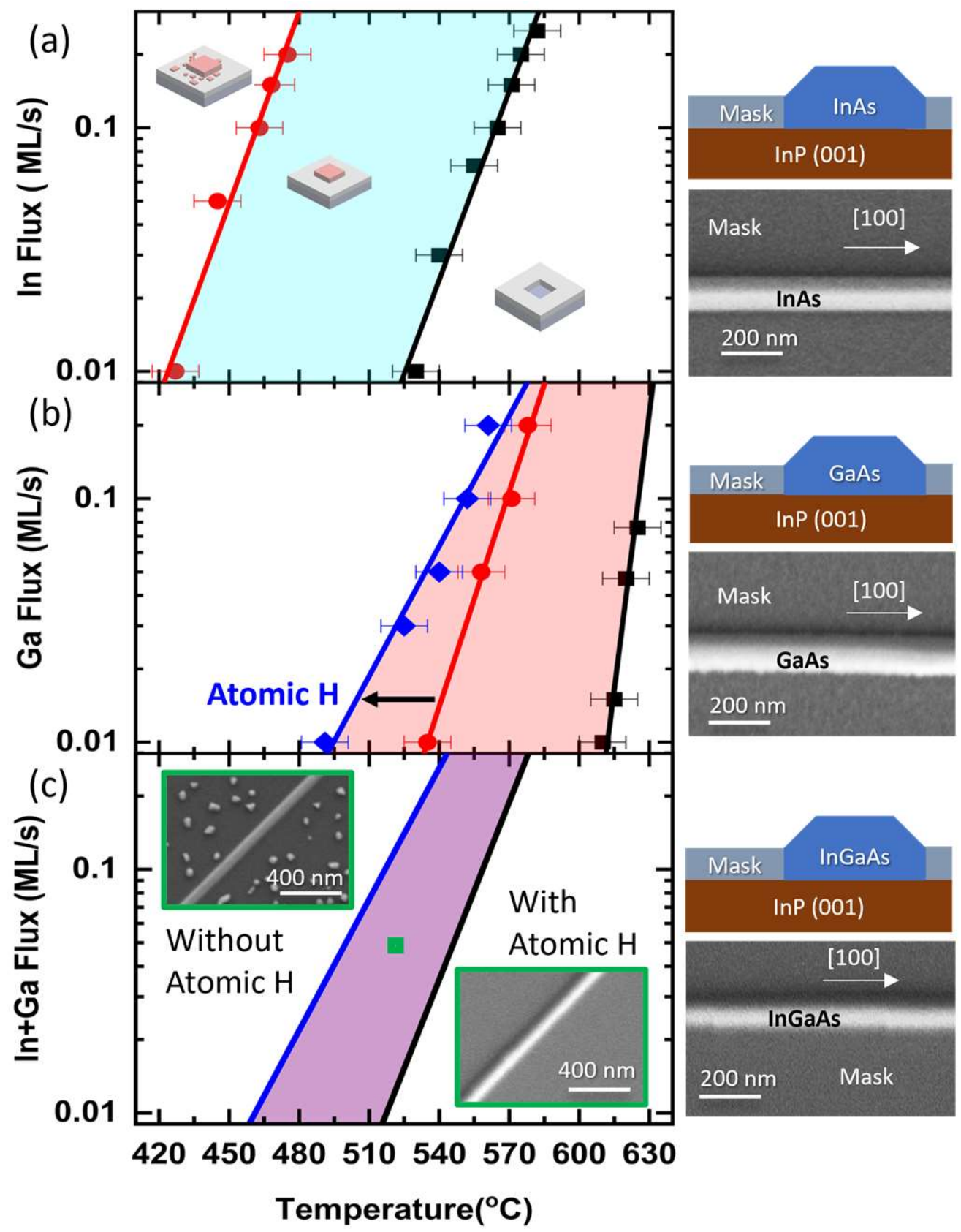

FIG. 1. Selectivity mapping for SAG of III-V semiconductors. (a) InAs selectivity map: The red and black lines denote the lower and upper bounds of the selectivity window. (b) GaAs selectivity map: The red and black lines denote the lower and upper bounds of the selectivity window. The use of atomic H shifts the lower bound further left (blue line), thereby widening the selectivity window. (c) The $In_{0.75}Ga_{0.25}As$ selectivity map is derived from data points in (a) and (b). The insets in (c) show SEM images of $In_{0.75}Ga_{0.25}As$ nanowires grown with and without atomic H grown at the point indicted by the green square. On the right of each selectivity map is a schematic and an SEM image of the associated grown nanowires demonstrating good selectivity.

## B. SAG InAs nanowires with/without buffer and/or capping layers

Processing of patterned SAG substrates begins with the deposition of a mask layer on the substrate. Nanowire networks are then lithographically patterned and transferred to the substrate by etching the mask layer. We deposited a 5 nm layer of aluminum oxide ($Al_2O_3$) through thermal atomic layer deposition, followed by the deposition of 15 nm of $SiN_x$ using plasma-enhanced chemical vapor deposition. The $Al_2O_3$ layer protects the surface of the InP(001) substrate from plasma damage and also acts as an etch stop layer during etching of $SiN_x$ using reactive ion etching. Before patterning the nanowires, micron-sized alignment marks are defined by a lift-off process using e-beam lithography followed by e-beam evaporation of 10 nm of chromium (Cr) and 90 nm of tungsten (W). These alignment marks are crucial for accurately locating the nanowires during multilayer lithography in device fabrication. The nanowire patterns are then defined using electron beam lithography and formed on the substrate through the dry etching of the $SiN_x$ layer and the wet etching of the $Al_2O_3$ layer using Transene D Al etchant. Finally, the prepared wafer is diced into 5 × 10 mm pieces and cleaned using UV ozone treatment and diluted HCl to remove any residual organic resists.

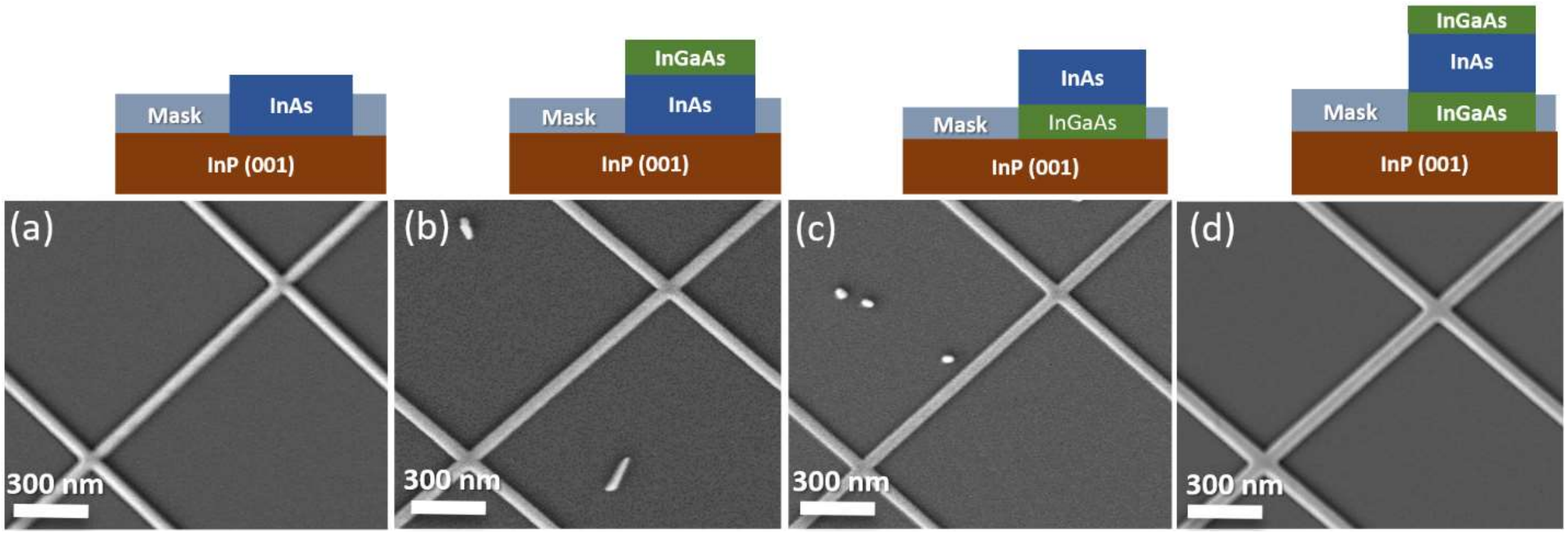

FIG. 2. SEM images of four sets of grown samples with schematic structure in the inset. (a) in-plane InAs nanowire on InP(001), (b) in-plane InAs nanowire with an InGaAs capping layer, (c) in-plane InAs nanowire with an InGaAs buffer layer, (d) in-plane InAs nanowire with both an InGaAs buffer and capping layer. The main nanowire is oriented along <100>.

The prepared die was outgassed at 250°C in an ultra-high vacuum chamber and transferred to the III-V MBE chamber. Native oxides were removed by soft thermal annealing at approximately 450°C under $As_4$ flux combined with atomic H. This method of oxide desorption, which occurs at a relatively lower temperature due to the presence of atomic H, produces an atomically smooth surface while avoiding void formation often seen with conventional high-temperature thermal desorption [35,37,38]. The temperature was monitored using a pyrometer, which had been calibrated based on the RHEED transition temperature of InAs [39], on a bare InP(001) reference wafer that was mounted along with the patterned die.

To study the role of closely lattice-matched buffer and capping layers in enhancing electrical properties, we fabricated the mask in a Hall bar geometry with a channel length of 2200 nm and a width of 120 nm along the <100> orientation. We prepared four sets of samples: (a) 60-nm-thick InAs, (b) 60-nm-thick InAs with a 15-nm-thick $In_{0.75}Ga_{0.25}As$ capping layer, (c) 60-nm-thick InAs with a 30-nm-thick $In_{0.75}Ga_{0.25}As$ buffer layer, and (d) 60-nm-thick InAs with a 30-nm-thick $In_{0.75}Ga_{0.25}As$ buffer layer and a 15-nm-thick $In_{0.75}Ga_{0.25}As$ capping layer, on InP(001) substrates. All four samples were grown at 520ºC at a growth rate of 0.08 ML/s for both InAs and $In_{0.75}Ga_{0.25}As$. $In_{0.75}Ga_{0.25}As$ buffer and capping layers were grown in the presence of atomic H flux of $5\times10^6$ Torr, while InAs was grown without atomic H. The nanowire thickness was confirmed by analyzing the height profile obtained from atomic force microscopy measurements, as shown in supplemental material Fig. S2. The SEM images of these Hall geometry nanowires in Fig. 2 show growth with a high selectivity, smooth faceting, and well-defined morphology of the nanowires and their junctions.

### C. Enhanced electrical properties

To form ohmic contacts when fabricating Hall bar devices on the four sets of samples, a 20-nm-thick aluminum (Al) layer was grown *in situ* using MBE with active cooling of the substrate holder using liquid nitrogen, after the growth of SAG nanowires. The Al layer was selectively etched for the Hall bar geometry using e-beam lithography, as illustrated in the inset of Fig. 3(c). This *in-situ* Al growth provides transparent interfaces between Al and III-V nanowire with minimal disorder, compared to other methods including *ex-* *situ* metal deposition after removing native oxide by Ar milling, wet-etching with HF solution, or passivation with ammonium sulfide [40]. Ohmic contacts were obtained in both Al/InAs and Al/ $In_{0.75}Ga_{0.25}As$ (10 nm)/InAs nanowires, as reported in Al/InAs 2DEG hybrid systems [22,41].

Electrical transport measurements were conducted at 2 K using a standard four-terminal current-bias setup, utilizing low-frequency lock-in amplifiers. Out-of-plane magnetic field was swept from -1T to 1T, and both Hall resistance ($R_{xy}$) and longitudinal resistance ($R_{xx}$) were measured simultaneously (Fig. 3). All four samples exhibit a linear dependence of Hall resistance with respect to the magnetic field. Electron density ($n_e$), electron mobility ($\mu_e$), and elastic mean free path ($l_e$) were extracted from the electrical transport measurements and summarized in Table I. The electron mobility of the InAs nanowires significantly increases with the incorporation of the $In_{0.75}Ga_{0.25}As$ buffer and/or capping layers. Specifically, the electron mobility of the InAs nanowire with both buffer and capping layers (5495 $cm^2/Vs$) is nearly three times higher than that of InAs nanowire without buffer and capping layers (1935 $cm^2/Vs$). Additionally, the mean free path of the InAs nanowire with both buffer and capping layers (109 nm) is more than twice that of the InAs nanowire without buffer and capping layers (43 nm), indicating fewer scattering sites in the InAs nanowire with both buffer and capping layers. This significant increase in electron mobility and mean free path is consistent with a reduction in interfacial defects and disorder due to the addition of closely lattice-matched InGaAs buffer and capping layers.

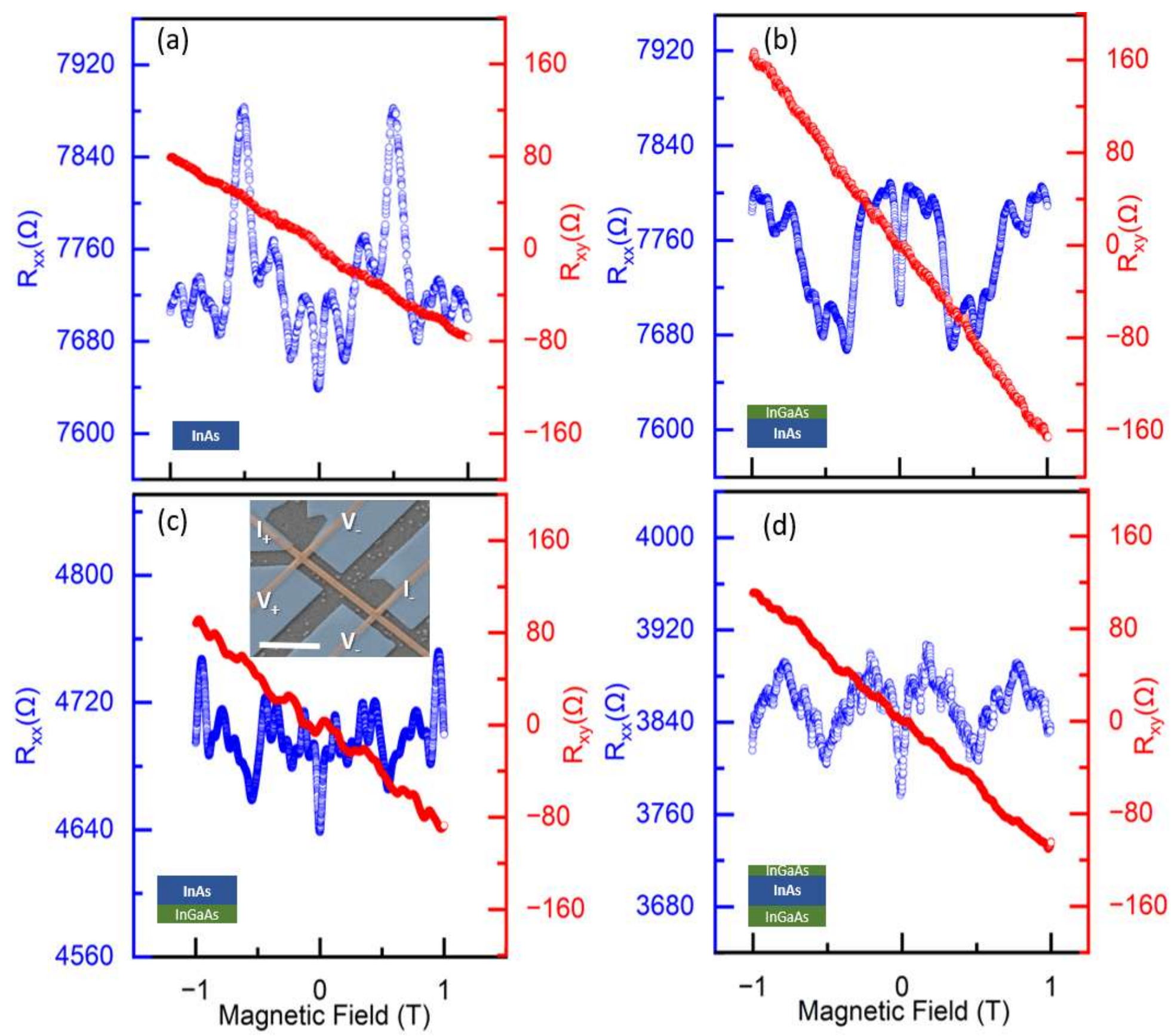

FIG. 3. Hall measurement data of InAs nanowires carried out at 2 K. Longitudinal magnetoresistance (blue curves) and Hall resistance (red curves) were measured for (a) SAG InAs, (b) SAG InAs with an InGaAs capping layer, (c) SAG InAs with an InGaAs buffer layer, and (d) SAG InAs with both an InGaAs as both buffer and capping layer. The nanowire schematics are shown on the bottom left. An SEM image of a representative Hall bar device is shown in the top inset of (c) in false color (Hall bar geometry nanowire in orange and aluminum contact pads in blue). The scale bar is 2μm.

TABLE I. Carrier density ($n_e$), mobility ($\mu_e$) and elastic mean free path ($l_e$) extracted from the Hall measurements of InAs NWs with/without buffer and/or capping layer. Spin-orbit length ($l_{so}$) and coherence length ($l_\emptyset$) were extracted from fitting Eq. (1) on WAL corrections to conductivity.

| Device | Channel | Buffer layer | Capping layer | $n_e$ ($10^{18}$ cm$^{-3}$) | $\mu_e$ (cm$^2$/Vs) | $l_e$ (nm) | $l_{so}$ (nm) | $l_\emptyset$ (nm) |
|---|---|---|---|---|---|---|---|---|
| (a) | InAs 60 nm | - | - | 1.30 | 1935 | 43 | 138 | 272 |
| (b) | InAs 60 nm | - | $In_{0.75}Ga_{0.25}As$ 15 nm | 0.63 | 3932 | 69 | 156 | 311 |
| (c) | InAs 60 nm | $In_{0.75}Ga_{0.25}As$ 30 nm | - | 1.20 | 3445 | 74 | 181 | 397 |
| (d) | InAs 60 nm | $In_{0.75}Ga_{0.25}As$ 30 nm | $In_{0.75}Ga_{0.25}As$ 15 nm | 0.92 | 5495 | 109 | 101 | 567 |

Furthermore, the longitudinal magnetoresistance for each sample reveals reproducible yet random conductance fluctuations, known as universal conductance fluctuations (UCF). These fluctuations arise from quantum interference effects when the carrier diffusion length surpasses the dimensions of the nanowires [14,42,43]. In conjunction with UCF, we observe a weak-antilocalization (WAL) dip around B = 0 T. This WAL feature signifies quantum interference combined with strong spin-orbit interaction on InAs nanowires [42,44]. The correction to the conductivity by WAL can be explained by quasi-classical theory as [45]:

$$\Delta G = -\frac{e^2}{h}\frac{1}{L}\left[3\left(\frac{1}{l_\emptyset^{\,2}} + \frac{1}{3l_{so}^{\,2}} + \frac{1}{l_B^{\,2}}\right)^{-\frac{1}{2}} - \left(\frac{1}{l_\emptyset^{\,2}} + \frac{1}{l_B^{\,2}}\right)^{-\frac{1}{2}} - 3\left(\frac{1}{l_\emptyset^{\,2}} + \frac{1}{3l_{so}^{\,2}} + \frac{3}{l_e^{\,2}} + \frac{1}{l_B^{\,2}}\right)^{-\frac{1}{2}} + \left(\frac{1}{l_\emptyset^{\,2}} + \frac{3}{l_e^{\,2}} + \frac{1}{l_B^{\,2}}\right)^{-\frac{1}{2}}\right], \quad (1)$$

where L is the channel length of the nanowire, $l_\emptyset$ is a coherence length, $l_{so}$ is spin-orbit length and $l_B$ is the magnetic dephasing length. Since the mean free path of the samples ranges from 40 nm to 110 nm which is smaller than the channel width(W), we are in the "dirty metal" regime [43,46]. In this regime magnetic dephasing length is given by $l_B^{\,2} = \frac{3l_m^{\,4}}{W^2}$ where $l_m = \sqrt{\frac{\hbar}{eB}}$ is the magnetic length. The change in conductivity, $\Delta G = G(B=0) - G(B)$ was fitted with the WAL correction given by Eq. (1) and is represented by the dashed lines in Fig. 4 in the magnetic field range of -90 mT to 90 mT. The extracted $l_\emptyset$ and $l_{so}$ values from the fit are summarized in Table I. The $l_{so}$ values range from 100 to 180 nm, which is in good agreement with earlier reports [43,47,48]. Although the $l_{so}$ value for the sample with both the buffer and capping layer is the lowest, indicating the strongest spin-orbit interaction, the $l_{so}$ values for the samples with only the capping or buffer layer are higher than those of the InAs nanowire alone. Therefore, no clear contribution of the buffer and capping layers to the spin-orbit interaction in InAs nanowires is observed. However, the coherence length increases with the addition of buffer and capping layers, and it is more than twice as large for the sample with both layers compared to that of InAs alone, confirming that the addition of buffer and capping layers enhances the electrical properties of the nanowire.

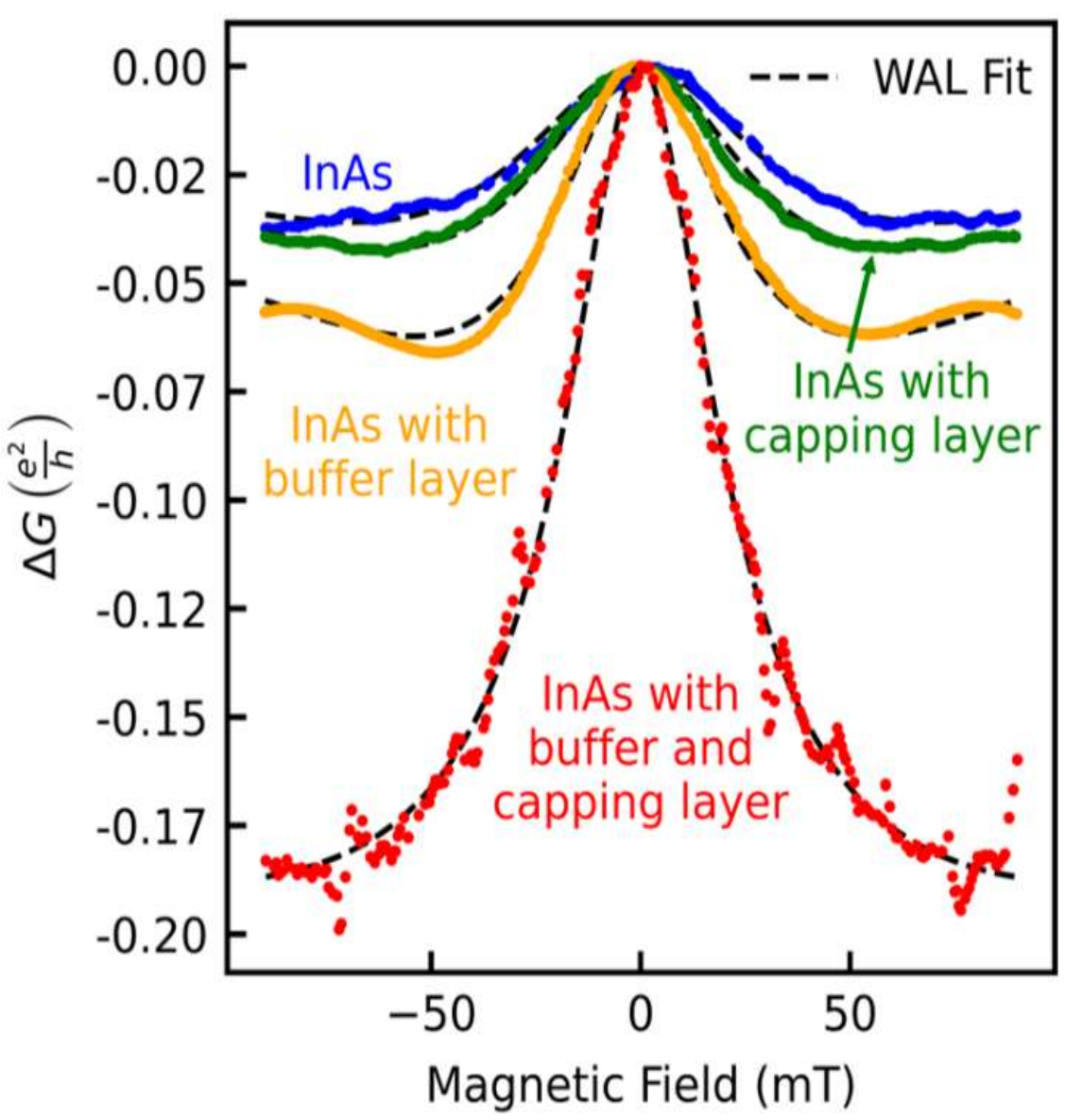

FIG. 4. Quantum corrections to conductivity with WAL for InAs (blue), InAs with a capping layer (green), InAs with a buffer layer (yellow), and InAs with both buffer and capping layers (red). Dashed lines represent fits using Eq. (1).

## 3. Conclusion

In this work, we investigated the SAG of in-plane InAs nanowires on $SiN_x$-mask-patterned InP(001) substrates and demonstrated that incorporating $In_{0.75}Ga_{0.25}As$ buffer (30 nm) and capping (15 nm) layers significantly enhances their electrical properties. By utilizing atomic H, we expanded the selectivity window for $In_{0.75}Ga_{0.25}As$, enabling its optimized growth as buffer and capping layers for InAs nanowires. Electrical transport measurements revealed that these layers effectively suppress interfacial defects and surface scattering, leading to a nearly threefold increase in electron mobility and a twofold enhancement in phase coherence length. These findings underscore the critical role of closely lattice-matched buffer and capping layers in improving the quality of SAG InAs nanowires, paving the way for their integration into high-performance electronics, optoelectronics, and quantum devices.

**Acknowledgments**

Material growth efforts of this work were supported by the U.S. Department of Energy, Office of Science, Basic Energy Sciences, Materials Sciences and Engineering Division. Characterization and measurement activities were supported by the Science Alliance at the University of Tennessee, Knoxville, through the Support for Affiliated Research Teams (StART) program. Substrate and device fabrication were performed as part of a user project at the Center for Nanophase Materials Sciences (CNMS), a U.S. Department of Energy, Office of Science User Facility at Oak Ridge National Laboratory.